\documentclass[fleqn,twoside]{article}
\usepackage{espcrc2}
\usepackage{epsfig}

\title{The Consequences of Non-Normality}

\author{I. Hip\address[Wup]{Department of Physics, University of
    Wuppertal, D-42097 Wuppertal, Germany}\thanks{Email:
    hip@theorie.physik.uni-wuppertal.de}, Th. Lippert\addressmark, H.
  Neff\address{NIC, c/o Research Center J{\"u}lich, D-52425 J{\"u}lich
    and DESY, D-22603 Hamburg, Germany}, K. Schilling\addressmark[Wup]
  and W. Schroers\addressmark[Wup]}

\begin{document}
 
\begin{abstract}
  The non-normality of Wilson-type lattice Dirac operators has
  important consequences --- the application of the usual concepts
  from the textbook (hermitian) quantum mechanics should be
  reconsidered.  This includes an appropriate definition of
  observables and the refinement of computational tools.  We show that the
  truncated singular value expansion is the optimal approximation to
  the inverse operator $D^{-1}$ and we prove that due to the
  $\gamma_5$-hermiticity it is equivalent to $\gamma_5$ times the
  truncated eigenmode expansion of the hermitian Wilson-Dirac
  operator. \vspace{1pc}
\end{abstract}

\maketitle
 
\section{INTRODUCTION}
\label{sec:introduction}
Wilson-type lattice Dirac operators are non-normal, i.e.
\begin{equation}
  [D, D^\dagger] \ne 0\,,
\end{equation}
and this fact has several important consequences. The essential
difference to normal operators is that they are not diagonalizable by
a unitary transformation. Instead they are diagonalizable (if at all)
by a similarity transformation of the form
\begin{equation}
  \mbox{diag}(\lambda_1, \ldots , \lambda_N) = X^{-1} \, D \, X\,,
\end{equation}
with $N = \mbox{dim}(D)$. It follows that for a given eigenvalue
$\lambda_i$ there are distinct left $\langle L_i |$ and right $| R_i
\rangle$ eigenvectors
\begin{equation}
  D | R_i \rangle = \lambda_i | R_i \rangle\,,
\end{equation}
\begin{equation}
  \langle L_i | D = \lambda_i \langle L_i |\,,
\end{equation}
in the sense that (in general) $\langle L_i | \ne (| R_i
\rangle)^\dagger$.

This fact must be taken into account when performing the spectral
decomposition
\begin{equation}
  D = \sum_{i = 1}^N \lambda_i | R_i \rangle \langle L_i |
  \ne \sum_{i = 1}^N \lambda_i |R_i \rangle \langle R_i |\,,
\end{equation}
but also raises the question of an appropriate generalization of the
other concepts familiar from the textbook (hermitian) quantum
mechanics. For example, if an observable in the continuum is defined
as $\langle \psi_i | A | \psi_i \rangle$, where $| \psi_i \rangle$ is
an eigenstate of the continuum operator $D$ --- what is the proper
generalization for the case where the left and the right eigenvectors
are distinct?

In \cite{Hip:2001hc} we advocate the use of $\langle L_i | A | R_i
\rangle$ instead of $\langle R_i | A | R_i \rangle$ as a kind of
improved observable. Note that in lattice field theory this question
is not of importance in the continuum limit, because the difference
between the two definitions disappears. However, this question was
also raised in the domain of solid state physics --- see
\cite{Hatano:1998} for an interesting discussion.

\section{TRUNCATED EIGENMODE EXPANSION}
\label{sec:trunc-eigenm-expans}
In this contribution we want to examine a different problem, which is
also connected to the non-normality of the Wilson-type lattice Dirac
operators.  The idea to approximate the inverse operator $D^{-1}$ by
just a few eigenmodes with eigenvalues close to zero is physically
appealing.  One hopes that the lowest modes dominate physics and that
the truncated eigenmode expansion
\begin{equation}
  D^{-1}_k = \sum_{i = 1}^{k \ll N} \frac{1}{\lambda_i}
  | R_i \rangle \langle L_i |
  \label{eq:lr-expansion}
\end{equation}
with the ordering
\begin{equation}
  \frac{1}{|\lambda_1|} \ge \frac{1}{|\lambda_2|} \ge \ldots \ge
  \frac{1}{|\lambda_k|} \ge \ldots \ge \frac{1}{|\lambda_N|}
\end{equation}
will be a fair approximation of the proper inverse $D^{-1} =
D^{-1}_N$.

\begin{figure}[tb]
  \epsfig{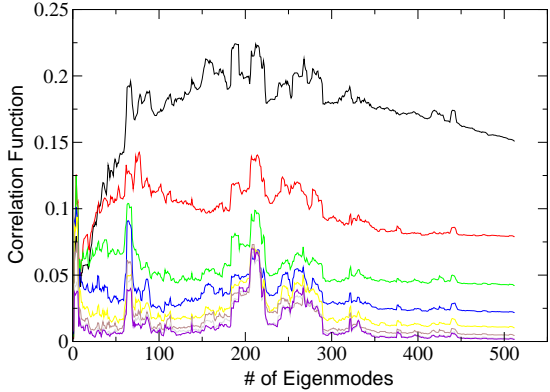}
  \vspace*{-5mm}
  \caption{Correlation function for several values of $\Delta t$
    ($\Delta t$ increases as one steps down from the top curve)
    plotted versus the number of eigenmodes used to approximate the
    inverse of the Wilson-Dirac operator --- for a given $k$, the
    correlation function is computed from $D_k^{-1}$, as defined in
    (\ref{eq:lr-expansion}). The dimension of $D$ is 512, so the final
    points give the values for the exact inverse of $D^{-1}$.
    \label{fig:b2-nhlr}}
\end{figure}

However, in \cite{Neff:2001zr} it was found that such an expansion is
unstable and that it does not uniformly saturate by increasing
$k$.  This is illustrated in Fig.~\ref{fig:b2-nhlr} where the values
of the correlation function for different temporal distances $\Delta
t$ are plotted versus $k$. (The illustration is just for two
dimensional QED on a $16^2$ lattice for $\beta = 2$, but it should be
stressed that for the discussion which follows this is irrelevant ---
the arguments are pure linear algebra and they will be valid for any
physical system and parameter set.)

By a closer examination of pseudospectra (for more details about
pseudospectra see \cite{www-oxford}) of $D$, we identified that the
spikes in Fig.~\ref{fig:b2-nhlr} correspond essentially to the
instabilities due to the non-normality of $D$. For non-normal
operators the truncated eigenmode expansion in the original
biorthogonal basis seems to be problematic. However, one can do better
than that.

\section{SINGULAR VALUE DECOMPOSITION}
\label{sec:sing-value-decomp}
For any matrix A there exists a so-called singular value decomposition
(SVD)
\begin{equation}
  A = U \, \Sigma \, V^\dagger = \sum_i \sigma_i | u_i \rangle \langle
  v_i |,
  \label{eq:svd}
\end{equation}
where $U$ and $V$ are unitary and $\Sigma$ is the diagonal matrix with
positive or zero entries on the diagonal --- the ``singular values''
$\sigma_i$.  We assume the following ordering:
\begin{equation}
  \sigma_1 \ge \sigma_2 \ge \ldots \sigma_N \ge 0\,.
\end{equation}
Among other interesting properties, one can prove the following
theorem (Th. 2.5.3. in \cite{Golub:1996bo}): 
\smallskip

\noindent If $k < r = \mbox{rank}(A)$ and
\begin{equation}
  A_k = \sum_{i = 1}^k \sigma_i | u_i \rangle \langle v_i |
\end{equation}
then
\begin{equation}
  \min_{\mbox{\scriptsize rank}(B)=k} ||A - B||_2 =
  ||A - A_k||_2 = \sigma_{k + 1}\,,
\end{equation}
i.e., $A_k$ is the closest matrix to $A$ that has rank $k$.  In other
words, the truncation of SVD is the optimal approximation to the
original matrix for a given $k$.

\begin{figure}[tb]
  \epsfig{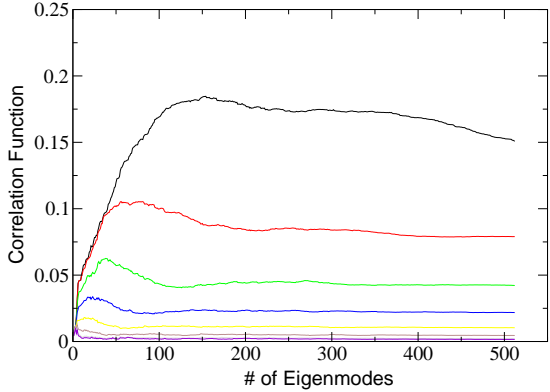}
  \vspace*{-5mm}
  \caption{Correlation function for several temporal distances $\Delta t$
    plotted versus the truncated singular value expansion, as defined
    by (\ref{eq:svd-expansion}). This time, the number of eigenmodes
    refers to the modes of the hermitian Wilson-Dirac operator $H$
    used to construct $H_k^{-1}$.
    \label{fig:b2-svd}}
\end{figure}

Actually, the drawback is that for a general matrix $A$ one needs two
sets of vectors: $|u \rangle$-s and $|v \rangle$-s. However, the
situation is better for the Wilson-type lattice Dirac operators (and
their inverses). Using the $\gamma_5$-hermiticity property
\begin{equation}
  D^\dagger = \gamma_5 D \gamma_5  \label{eq:g5-hermiticity}
\end{equation}
we are going to prove that the truncated SVD of $D^{-1}$ is identical
to
\begin{equation}
  D^{-1}_k = \gamma_5 \sum_{i = 1}^k \frac{1}{\mu_i}
  | w_i \rangle \langle w_i | = \gamma_5 H^{-1}_k, 
  \label{eq:svd-expansion}
\end{equation}
with the ordering according to the absolute values of inverse $\mu$-s:
\begin{equation}
  \frac{1}{|\mu_1|} \ge \frac{1}{|\mu_2|} \ge \ldots \ge
  \frac{1}{|\mu_k|} \ge \ldots \frac{1}{|\mu_N|}\,.
\end{equation}
The $\mu$-s and $|w \rangle$-s are the eigenvalues and eigenvectors of
the hermitian Wilson-Dirac operator
\begin{equation}
  H = \gamma_5 D = W M W^\dagger =
  \sum_{i = 1}^{N} \mu_i | w_i \rangle \langle w_i |\,.
\end{equation}
The hermiticity of $H$ follows from the $\gamma_5$-hermiticity
property (\ref{eq:g5-hermiticity}). Thus $H$ is diagonalizable by a
unitary transformation $W$ and has real eigenvalues $\mu_i$, ordered
on the diagonal of the diagonal matrix $M$.

Per construction, $D^\dagger D$ is a hermitian matrix and it is easy
to see that
\begin{equation}
  D^\dagger D = W M W^\dagger W M W^\dagger = W M^2 W^\dagger.
\end{equation}
However, by using the singular value decomposition (\ref{eq:svd}) of
$D$ it follows that
\begin{equation}
  D^\dagger D = V \Sigma U^\dagger U \Sigma V^\dagger 
  = V \Sigma^2 V^\dagger.
\end{equation}
As $V$ is unitary and $\Sigma^2$ diagonal, one must have (with
appropriate ordering)
\begin{equation}
  \sigma_i^2 = \mu_i^2 \quad \Rightarrow \quad \sigma_i = | \mu_i |\,,
\end{equation}
i.e., the singular values of $D$ are just the absolute values of the
eigenvalues of the hermitian Wilson-Dirac operator $H$.  Also, the $|v
\rangle$-s should correspond to the $|w \rangle$-s (there is an
arbitrary phase factor, but it cancels in (\ref{eq:svd-expansion})).
Finally, we can rewrite:
\begin{equation}
  D = U \Sigma V^\dagger = U \mbox{sign}(M) M W^\dagger 
  = \gamma_5 W M W^\dagger
\end{equation}
which makes it obvious that
\begin{equation}
  U = \gamma_5 W \mbox{sign}(M) \; \Rightarrow \;
  | u_i \rangle = \mbox{sign}(\mu_i)\, \gamma_5 | w_i \rangle.
\end{equation}
Hence, the truncated singular value expansion of $D$ is
{\arraycolsep=1pt
  \begin{eqnarray}
    D_k & = & \sum_{i = 1}^k \sigma_i | u_i \rangle \langle v_i | 
    = \sum_{i = 1}^k \sigma_i\, \mbox{sign}(\mu_i)\,
    \gamma_5 | w_i \rangle \langle w_i | \nonumber \\
    & = & \gamma_5 \sum_{i = 1}^k \mu_i | w_i \rangle \langle w_i |
    = \gamma_5 H_k\,.
  \end{eqnarray}}
As $D^{-1}$ also satisfies the $\gamma_5$-hermiticity property
(\ref{eq:g5-hermiticity}), the same reasoning applies and
(\ref{eq:svd-expansion}) follows.

\section{CONCLUSION}
\label{sec:conclusion}
Fig.~\ref{fig:b2-svd} illustrates that, as expected, by using the
truncated singular value expansion, non-normal artefacts disappear and
the saturation as a function of $k$ is superior to the expansion in
the eigenmodes of $D$. This approach has already been successfully
used in a realistic QCD setting \cite{Neff:2001zr}.

\vspace*{3mm}

\noindent {\bf Acknowledgments:} I.~H.~thanks the DOE's Institute of
Nuclear Theory at the University of Washington for its hospitality and
the DOE for partial support during the completion of this work.
W.~S.~is supported by the DFG Graduiertenkolleg ``Feldtheoretische
Methoden in der Elemen\-tar\-teil\-chen\-theorie und Statistischen
Physik''.

\end{document}